\documentstyle[aas2pp4]{article} 

\input epsf.tex

\newcommand {\Mpc}   {\mbox{$h^{-1}$ Mpc \,}}  
  
\newcommand{\mincir}{\raise -2.truept\hbox{\rlap{\hbox{$\sim$}}\raise5.truept  
\hbox{$?$}\ }}  
\newcommand{\gr}{\kern 2pt\hbox{}^\circ{\kern -2pt K}} 
\newcommand{\magcir}{\raise -2.truept\hbox{\rlap{\hbox{$\sim$}}\raise5.truept  
\hbox{$?$}\ }}

\begin{document}

\title{Best-fit parameters of MDM model from Abell-ACO power spectra and 
mass function}  
  
\author{B.~Novosyadlyj}

\affil{Astronomical Observatory of L'viv State University, Kyryla and  
Mephodia str.8, 290005, L'viv, Ukraine}

\begin{abstract} 
 
The possibility of determining MDM model parameters on the basis 
of observable data on the Abell-ACO power spectrum and mass function is 
analysed. It is shown that spectrum area corresponding to these data 
is sensitive enough to such MDM model parameters as neutrino mass 
$m_{\nu}$, number species of massive neutrino $N_{\nu}$, baryon content 
$\Omega_b$ and Hubble constant $h\equiv H_0/100km/s/Mpc$. The $\chi^2$ 
minimization method was used for their determination. If all these 
parameters are under searching then observable data on the Abell-ACO 
power spectrum and mass function prefer models which have parameters 
in the range  
$\Omega_{\nu}$ ($\sim 0.4-0.5$), low $\Omega_b$ ($\le 0.01$) and $h$ 
($\sim 0.4-0.6$). The best-fit parameters are as follows: 
$N_{\nu}=3$, $m_{\nu}=4.4eV$, $h=0.56$, $\Omega_b\le 0.01$. 
The high-$\Omega_b\sim 0.4-0.5$ solutions are obtained when mass of neutrino 
is fixed and $\le 3eV$. 
 
To explain the observable excessive power at $k\approx 0.05h/Mpc$ the 
peak of Gaussian form was introduced in primordial power spectrum. Its 
parameters (amplitude, position and width) were determined along with 
the MDM model parameters. It decreases $\chi^2$, increases the bulk
motions, but  does not change essentially the best-fit MDM parameters. 
 
It is shown also that models with the
median $\Omega_{\nu}\sim 0.2-0.3$ ($m_{\nu}\sim 2.5$, $N_{\nu}\sim 2-3$) 
and $\Omega_b=0.024/h^2$, which match constraints arising
from cosmological nucleosynthesis and high redshift objects, are not
ruled out by these data ($\Delta \chi^2<1$).    

\end{abstract}

 
\section{Introduction} 
 
The observable data on large scale structure of the Universe obtained 
during last years and coming from current experiments and observational 
program give a possibility to determine  more exactly the parameters of 
cosmological models and the nature of the dark matter. Up till now  the most 
certain data are about the largest scale inhomogeneities of the current particle 
horizon of the order of $\sim 7000\Mpc$ ($h\equiv H_0/100\;km/s/Mpc$, $H_0$ is today 
Hubble constant) which are obtained from the study of all-sky temperature 
fluctuations of cosmic microwave background (CMB) with $\sim 10^{o}$ angular 
resolution by the space experiment COBE (\cite{smo92,ben94,ben96}). 
According to them the  primordial power spectrum 
of density fluctuations is approximately scale invariant $P_{pi}(k)=Ak^{n}$ 
with $n=1.1\pm 0.2$ that well agrees with the predictions of standard inflation 
model of the Early Universe ($n=1$, $\Omega_0=1$). 
Besides, they most certainly determine 
the amplitude of a linear power spectrum (or normalization constant $A$) 
which does not depend on any transition processes, nonlinearity effects and 
other phenomena connected with the last stages of large scale structure formation. 
On the contrary, 
the CMB temperature fluctuations at degree and sub-degree scales as well as 
the space distributions of the cluster of galaxies, galaxies, quasars, Lyman-$\alpha$ 
clouds, etc. are defined by those processes and also depend essentially on 
the nature of the dark matter. Theoretically it is taken into account by introducing 
the transfer function $T(k)$ which transforms the primordial (post-inflation 
spectrum) into the postrecombination (initial) one - $P(k)=P_{pi}(k)T^{2}(k)$, which 
defines all characteristics of the large scale structure of the Universe. 
The transfer function depends also on the curvature of the Universe or the
present energy density in units of critical density, $\Omega_{0}$, vacuum energy 
density or cosmological constant $\Omega_{\Lambda}$, content of baryons 
$\Omega_b$, and values of the Hubble constant. 
The theory of a large scale structure formation is so far advanced  today 
that all these dependencies can be accurately calculated for the fixed model by public 
available codes (e.g. CMBfast one by \cite{sz96}). The actual problem now 
is the determination of the nature of the dark matter 
and the rest of the above mentioned parameters by means of comparison of theoretically 
predicted and observable characteristics of the large scale structure of 
the Universe. 
 
As most advanced candidates for the dark matter  are 
cold dark matter (CDM), particles like axions, hot dark matter (HDM), particles 
like massive neutrinos with $m_{\nu}\sim 1-20eV$ and baryon low luminosity 
compact objects. The last ones can not dominate as it results from the cosmological 
nucleosynthesis constraints ($\Omega_bh^{2}\le 0.024$, 
\cite{tyt96,son97,sct97}) and 
observation of microlensing events in the experiments like MACHO, DUO, etc. 
The pure HDM model conflicts with the existence of high redshift objects, the 
pure CDM one, on the contrary, overpredicts them. Therefore mixed dark matter 
model (CDM+HDM+baryons) with $\Omega_{HDM}\equiv \Omega_{\nu}\le 0.3$ looks more 
viable. The advantage  of these models is a small number of free parameters. 
But today it is  understood already that models with the minimal number of free 
parameters, such as a standard cold dark matter (sCDM, one parameter) or a standard 
cold plus 
hot mixed dark matter (sMDM, two parameters) only marginally match the 
observable data. A better agreement between theoretical predictions and 
observable data is achieved in the models with a larger number of free parameters 
(tilted CDM, open CDM, CDM or MDM with the cosmological term, see review in 
\cite{vkn98} and references therein). 
 
The oscillations of solar and atmospheric neutrinos registered 
by SuperKamiokande experiment show that the difference of rest masses between 
$\tau -$ and $\mu$-neutrinos is $0.02<\Delta m_{\tau \mu} < 0.08eV$
\cite{fu98,pr98}. It also gives a lower limit for the mass of neutrino $m_{\nu}\ge
|\Delta m|$ and does not exclude models with  cosmologically
significant values $\sim 1-20eV$. Therefore, at least two  species of
neutrinos can have approximately equal masses in this range.  Some
versions of elementary particle theories predict  $m_{\nu _e}\approx
m_{\nu _\tau}\approx 2.5eV$ and  $m_{\nu _{\mu}}\approx m_{\nu _s}\sim
10^{-5}eV$, where  ${\nu _e}$, ${\nu _\tau}$, ${\nu _{\mu}}$ and ${\nu
_s}$ denote the  electron, $\tau -$, $\mu -$  and sterile neutrinos
accordingly (e.g. \cite{dol95}).  The strongest 
upper limit for the neutrino mass comes from the data on a large scale structure of 
our Universe: $\sum_{i} m_{\nu_i}/93h^2\le 0.3$ 
(\cite{hol89,dav92,sch92,van92}, Novosyadlyj 1994, \cite{pog95,ma96,vkn98}),
that for $h=0.8$  (the upper observable limit for $h$) 
gives $\sum_{i} m_{\nu_i}\le 18eV$. It is interesting that the upper limit for the mass 
of electron neutrino obtained from supernova star burst SN1987A neutrino 
signal is approximately the same $m_{\nu_e}\le 20eV$. 
 
Is it possible to find the best fit neutrino mass 
from experimental data on a large 
scale structure of the Universe? The problem is that it must 
be determined together with 
other large number uncertain parameters such as $h$, $\Omega_m$, $\Omega_b$, 
etc. Here we study the possibility of finding them by $\chi ^2$ minimization 
method. Realization of such a task became possible in principle after 
the appearance in literature of accurate analytical approximations of transfer 
function for mixed dark matter model in at least 4-dimension space of 
the above mentioned cosmological parameters 
$T(k;\Omega_{b},m_{\nu},N_{\nu},h)$ (\cite{eh3,nov98}). 
 That is why that even CMBfast 
codes are too bulky and slow yet for searching the cosmological parameters 
by the methods of minimization of $\chi^2$, like Levenberg-Marquardt one 
(see \cite{nr92}). 
 
The next problem is a choice of the observable data suitable for the solution of 
this task. They must be enough accurate, sensitive to those parameters and 
not too dependent on the model assumptions about the formation and nature of 
objects. The most sensitive to the presence of neutrino component are scales of order 
and smaller of its free-streaming (or Jeans) scale 
$k\ge k_{J}(z)=8\left({m_{\nu}\over 10eV}\right)/\sqrt{1+z}\Mpc$ 
because  perturbations at these scales are suppressed and it is 
imprinted in the transfer function of the HDM  component. At $z\sim 0$ for 
cosmologically significant neutrino masses it is approximately galaxy clusters 
scale. The power spectrum reconstructed from space 
distributions of galaxies is distorted significantly by nonlinearity effects 
the accounting of which is model dependent (\cite{pd94}). The models of 
the formation of smaller 
scale structures or high redshift objects (e.g. Lyman-$\alpha$ damped systems, 
Lyman-$\alpha$ clouds, quasars etc.) contain the additional assumptions and 
parameters which makes their using rather problematic in such an approach. The CMB 
temperature anisotropy at subdegree angular scales (first and second acoustic 
peaks) has minimal additional assumptions (e.g. secondary ionization) 
but its sensibility to the presence of neutrino component is low ($\le 10\%$, 
\cite{dod95}). 
These data are sensitive and suitable for determination by $\chi^2$ 
minimization methods other set of parameters such as tilt of primordial 
spectrum $n$, $\Omega_0$, $h$, $\Omega_b$, $\Omega_{\Lambda}$ or/and parameters 
of scaling seed models of structure formation (see \cite{lin97,dur97}). 
 
The data on Abell-ACO power spectrum and function mass of rich clusters 
seem to be suitable for 
determining the best fit values of $m_{\nu}$ and $N_{\nu}$ because they 
do not depend on above mentioned additional assumptions. 
 
The data on rich clusters power spectrum (\cite{ein97}) were used by 
Eisenstein et al. 1997 and Atrio-Barandela  et al. 1997 for analyzing  $\sim
100\Mpc$ clustering.  The first collaboration group tried to explain the
narrow peak in the power  spectrum at $\sim 100\Mpc$ scale by baryonic
acoustic oscillations in low-  and high-$\Omega_{0}$ models
($\Omega_0=\Omega_{CDM}+\Omega_{b}$).  In both cases such an approach needs very
high  content of baryons $\Omega_{b}$ ($>0.3$), 	that is essentially
out of the cosmological  nucleosynthesis constraints. The second one has shown
that this feature  is in agreement with Saskatoon data (\cite{net97}) on
$\Delta T/T$ power spectrum at  subdegree angular scales. 
They have concluded that these data prefer 
models with built-in scale in the primordial power spectrum which can be 
generated in the more complicated inflation scenario (e.g. double one). 
 
For reducing the number of free parameters we restrict ourselves to analysis 
within the framework of the matter dominated Universe and standard inflation scenario: 
$\Omega_m=\Omega_0=1$, $n=1$ without the tensor mode of cosmological perturbations. 
The free parameters in our task will be baryon content $\Omega_b$, 
dimensionless Hubble constant $h$, neutrino mass $m_{\nu}$, and number 
species of neutrinos with equal masses $N_{\nu}$. 
 
The outline of this paper is as follows: the observable data which will be used 
here are 
described in Section 2. The method of determination of parameters and its 
testing  are described in Sect. 3. Results of best fit finding of parameters 
under different combination of free and fixed ones are presented in Sect. 4. 
Discussion of results and conclusions are given in Sect. 5 and 6 accordingly.

\section{Experimental data set} 
 
The most favorable data for the search of best fit cosmological parameters are 
real power spectrum reconstructed from redshift-space distribution of Abell- 
ACO clusters of galaxies (\cite{ein97,ret97}). It is biased linear spectrum 
reliably estimated for $0.03\le k\le 0.2h/Mpc$ whose position of maximum 
($k_{max}\approx 0.05h/Mpc$), inclination before and after it are sensitive 
to baryon content $\Omega_{b}$, Hubble constant $h$, neutrino mass $m_{\nu}$ 
and number species of massive neutrinos $N_{\nu}$ (see Fig.1-4). Here in 
numerical calculations the data of last estimation of power spectrum by 
\cite{ret97} will be used. All the sources of systematic and statistical 
uncertainties as well as window function and differences between Abell 
and ACO parts of sample have been accurately taken into account there. The 
values of the Abell-ACO power spectrum for 13 values of $k$
$\tilde P_{A+ACO}(k_j)$ ($j=1,13$) and their $1\sigma$ errors  
are presented in Table 1 and are shown in figures. 
 
Other observable data which will be used here are constraints of amplitude 
of the fluctuation power spectrum at cluster scale derived from cluster mass 
and X-ray temperature functions. It is usually formulated as a constraint for 
density fluctuations in top-hat sphere of 8\Mpc radius, $\sigma_{8}$, which 
can be easy calculated for the given initial power spectrum $P(k)$: 
$$\sigma_{8}^{2}={1\over 2\pi^{2}}\int_{0}^{\infty}k^{2}P(k;\Omega_{b},h,m_{\nu}, 
N_{\nu})W^{2}(8k/h)dk,\eqno(1)$$ 
where $W(x)=3(sin x-x cos x)/x^3$ is Fourier transformation of top-hat window 
function. The different collaboration groups gave similar results 
which are in the range of $\tilde \sigma_8\sim 0.5-0.7$. 
The new optical determination of 
the  mass function of nearby galaxy clusters (\cite{gir98}) gives 
median values: $\tilde \sigma_{8}=0.60\pm 0.04$\footnote{Jeans scale for 
neutrino component in all cases analysed 
here is smaller than the cluster scales therefore all the matter is clustered and 
the term $\Omega^{0.6}$ in the original form is omited}. 
It matches very well the cluster X-ray temperature function (\cite{via96}). 
For taking into account the data of other authors I shall be more conservative 
and will use it with $3\sigma$ error bars instead of $1\sigma$ one. But, as we 
will see, it does not rule out predicted $\sigma_{8}$ value from the $1\sigma$ 
limit of the observable one by \cite{gir98} for best fit parameters 
determined here. 
 
\begin{table}[th] 
\caption{Experimental data set.} 
\def\onerule{\noalign{\medskip\hrule\medskip}} 
\medskip 
\halign{#&\quad #&\quad #&\quad #&\quad #&\quad #&\quad #&\quad #&\quad # 
&\quad #&\quad #&\quad #&\quad #&\quad #&\quad #\cr 
\noalign{\hrule\medskip} 
  No &  $k_j$ &     $\tilde y_j$  &     $\Delta \tilde y_j$\cr 
\noalign{\hrule\medskip} 
  1 &  0.030    &$9.312\cdot 10^4$&   $\pm 59723.65$  \cr 
  2 &  0.035    &$1.037\cdot 10^5$&   $\pm 65488.2$   \cr 
  3 &  0.040    &$1.039\cdot 10^5$&   $\pm 58014.15$  \cr 
  4 &  0.047    &$1.258\cdot 10^5$&   $\pm 51005.75$  \cr 
  5 &  0.054    &$1.448\cdot 10^5$&   $\pm 68638.6$   \cr 
  6 &  0.062    &$1.016\cdot 10^5$&   $\pm 39184.6$   \cr 
  7 &  0.072    &$8.098\cdot 10^4$&   $\pm 25179.7$   \cr 
  8 &  0.083    &$5.444\cdot 10^4$&   $\pm 21925.45$  \cr 
  0 &  0.096    &$5.303\cdot 10^4$&   $\pm 24914.75$  \cr 
 10 &  0.11     &$3.853\cdot 10^4$&   $\pm 13344.5$   \cr 
 11 &  0.13     &$2.031\cdot 10^4$&   $\pm  8546.35$  \cr 
 12 &  0.15     &$2.039\cdot 10^4$&   $\pm  9804.3$   \cr 
 13 &  0.17     &$1.691\cdot 10^4$&   $\pm  9383.21$  \cr 
 14 & $\sigma_8$&  0.60           &   $\pm  0.12$     \cr 
\noalign{\hrule\medskip} 
} 
\end{table}

The COBE 4-year data will be used here for normalization of power spectra. 
A useful fit for them is the amplitude of density perturbation of the
horizon crossing scale $\delta_h$, which for a flat model with the $n=1$ equals 
$\delta_h=1.94\cdot 10^{-5}$ (\cite{lid96,bun97}). Taking into account the 
definition of $\delta_h$ (\cite{lid96}) and the power spectrum, the 
normalization constant $A$ is calculated as 
$$A=2\pi^{2}\delta_{h}^{2}(3000/h)^{4}Mpc^{4}.$$ 
 
\section{Method and its testing} 
 
The Abell-ACO power spectrum is connected with matter one by means of the
cluster biasing parameter $b_{cl}$: 
$$P_{A+ACO}(k)=b_{cl}^{2}P(k;\Omega_b,h,m_{\nu},N_{\nu}).\eqno(2)$$ 
For fixed parameters $\Omega_b$, $h$, $m_{\nu}$, $N_{\nu}$ and $b_{cl}$ 
the values of $P_{A+ACO}(k_j)$ are calculated for the same $k_j$ as in 
Table 1 and $\sigma_8$ according to (1). Let's denote them by $y_j$ 
($j=1,...,14$), where $y_{1},...,y_{13}$ correspond $P_{A+ACO}(k_1),..., 
P_{A+ACO}(k_{13})$, and $y_{14}$ is $\sigma_{8}$. Their deviation from 
observable data set (noted by the tilde) can be described by $\chi^2$: 
$$\chi^{2}=\sum_{j=1}^{14}\left({\tilde y_j-y_j \over \Delta \tilde y_j}
\right)^2,\eqno(3)$$ 
where $\tilde y_j$ and $\Delta \tilde y_j$ are experimental data set 
and their dispersion accordingly. Then parameters 
$\Omega_b$, $h$, $m_{\nu}$, $N_{\nu}$ and $b_{cl}$ or some part from 
them can be determined by minimizing $\chi^2$ using Levenberg-Marquard 
method (\cite{nr92}). The derivatives of predicted values on search 
parameters which are required by this method will be calculated numerically. 
The step for their calculation was experimentally assorted and is 
$10^{-5}$ of the values for all parameters. 

The analytical approximation of MDM transfer function 
will be used in the form: 
$$\begin{array}{l} 
T_{MDM}(k;\Omega_{b},h,m_{\nu},N_{\nu};z)=\\ 
\qquad {}T_{CDM+b}(k;\Omega_{b},h;z) 
D(k;\Omega_{b},h,\Omega_{\nu},N_{\nu};z), 
\end{array}$$ 
where $T_{CDM+b}(k;\Omega_{b},h;z)$ is the transfer function by \cite{eh2} for 
CDM+baryon system ($z$ is redshift), 
the correction factor for the HDM component $D(k)$ was used in the form given by
\cite{nov98}. It is correct in a sufficiently wide range of
search  parameters (for a more detailed analysis of its accuracy see in
\cite{nov98}).  We suppose the scale invariant primordial power spectrum
because  the initial power spectra of MDM models now is as follows: 
$P_{MDM}(k)=AkT_{MDM}^{2}(k;\Omega_{b},h,m_{\nu},N_{\nu};z)$.

The method was tested in the following way. I calculated the MDM 
power spectrum for the given parameters (e.g. $\Omega_b=0.15$, $\Omega_{\nu}=0.2$, 
$N_{\nu}=1$, $h=0.5$) using CMBfast code, normalized to 4-year COBE data, 
calculated $\tilde \sigma_8$ and interpolated $P(k)$ for the same $k_j$ 
($j=1,...,13$) which are in Table 1. Then I have took cluster biasing 
parameter $b_{cl}=3$ and calculated model $\tilde P_{A+ACO}(k_j)$.  The 
'experimental' errors for them as well as for $\tilde \sigma_8$ I have 
suggested to be the same as relative errors from Table 1. 
These model experimental data like the ones in Table 1 
were used for search of parameters $\Omega_b$, $h$, $\Omega_{\nu}$, 
and $b_{cl}$ ($N_{\nu}$ is fixed and the same).  The initial (or start) 
values of the parameters I have put as 
random deviated from the given ones. In all cases the code found all the given 
parameters with high accuracy.

\section{Dependence of density fluctuations power spectra at cluster scale on 
cosmological parameters} 
 
Before finding of the best-fit parameters let's look how the power 
spectrum of density fluctuations at cluster scale depends on 
search parameters. For this we leave only $b_{cl}$ as a free parameter 
and fix the remaining ones. In Fig.1  such a dependence of rich cluster power 
spectra on $\Omega_{\nu}$ 
is shown for $h=0.5$, $\Omega_{b}=0.05$ and $N_{\nu}=1$. 
The r.m.s. of density fluctuations in the top-hat sphere of $8\Mpc$ radius 
in models with $\Omega_{\nu}=$0.1, 0.2, 0.3, 0.4 are $\sigma_{8}=$ 
0.93, 0.81, 0.75, 0.71 accordingly. The best-fit values of $b_{cl}$ 
are presented in the caption of Fig.1. The deviations of the predicted rich cluster 
power spectra and mass function in these models from the observable ones are 
correspondingly 
$\chi^{2}=17.3,\;9.88,\;6.64,\;5.33$. Therefore, for the MDM model with 
$h=0.5$, $\Omega_{b}=0.05$ and $N_{\nu}=1$ Abell-ACO power spectrum and 
mass function prefer high $\Omega_{\nu}$ ($\sim 0.3-0.4$). 
\begin{figure}[th] 
\epsfxsize=9truecm 
\epsfbox{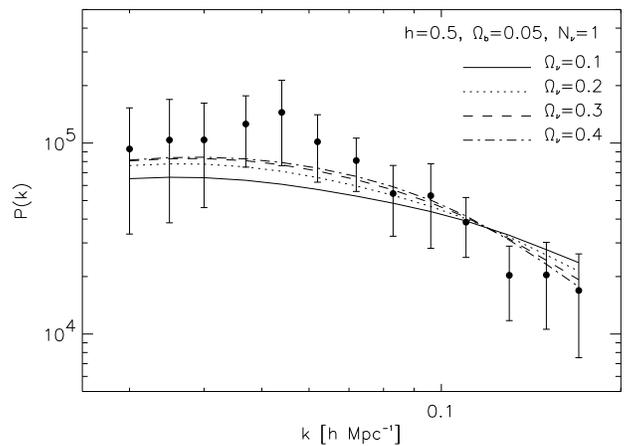} 
\caption{The rich cluster power spectrum for MDM models 
with different $\Omega_{\nu}$ ($N_{\nu}$, $\Omega_{b}$ and $h$ are fixed). 
The filed circles are experimental Abell-ACO power spectrum by Retzlaff 
et al. 1997. 
The best-fit biasing parameters for the models with $\Omega_{\nu}=$0.1, 0.2, 0.3, 
0.4 are $b_{cl}=$2.6, 2.8, 2.9, 2.9 accordingly.} 
\end{figure} 
 
Now we repeat the same calculations for different number species of 
massive neutrinos $N_{\nu}=1,\;2,\;3$ and fixed $\Omega_{\nu}=0.2$ (Fig.2). 
The $\sigma_{8}$'s for these 3 models are 0.81, 0.73, 0.68 accordingly, the 
corresponding deviations of predicted rich cluster power spectra and 
mass functions from the observable ones respectively are 
$\chi^{2}=$9.88, 6.48, 5.54. 
So, the MDM model with three species of equal mass neutrino is 
preferable. 
\begin{figure}[th] 
\epsfxsize=9truecm 
\epsfbox{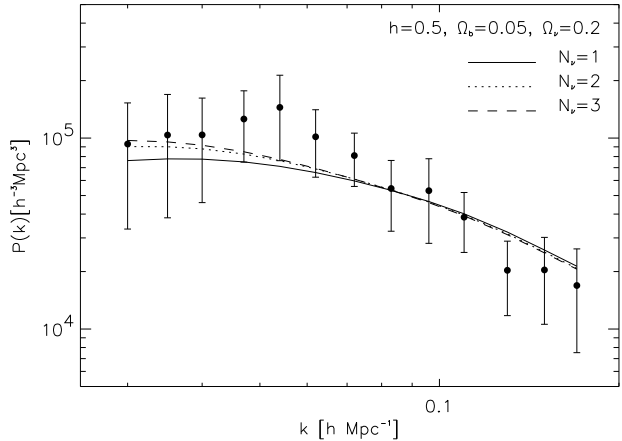} 
\caption{The  rich cluster power spectrum for MDM models 
with a varying number of species of massive neutrino $N_{\nu}$ 
($\Omega_{\nu}$, $\Omega_{b}$ and $h$ are fixed). 
The filed circles are the same as in Fig.1. 
The best-fit biasing parameters for models with $N_{\nu}=$1, 2, 3 are 
$b_{cl}=$2.8, 3.1, 3.3 accordingly.} 
\end{figure} 
 
In the first two cases ($h$ fixed and equal 0.5) the mass of neutrino was 
different for differing $\Omega_{\nu}$ ($N_{\nu}$ fixed) and $N_{\nu}$ 
($\Omega_{\nu}$ fixed) because they are connected by relations 
$$m_{\nu}=93\Omega_{\nu}h^{2}/N_{\nu}.\eqno(4)$$ 
Let's fix the neutrino 
mass ($m_{\nu}=2.5eV$), suggest that $N_{\nu}=2$ and repeat calculations for 
different $h=$0.5, 0.6, 0.7. The results are shown in Fig.3. $\sigma_{8}$ 
for these 3 models are following 0.71, 0.98, 1.24. 
The $\chi^2$ for all points of power spectrum and $\sigma_8$ 
are 5.72, 19.9 and 42.6 accordingly. Therefore, 
when neutrino mass is fixed (by laboratory experiments for example) 
the data prefer low $h$. 
\begin{figure}[th] 
\epsfxsize=9truecm 
\epsfbox{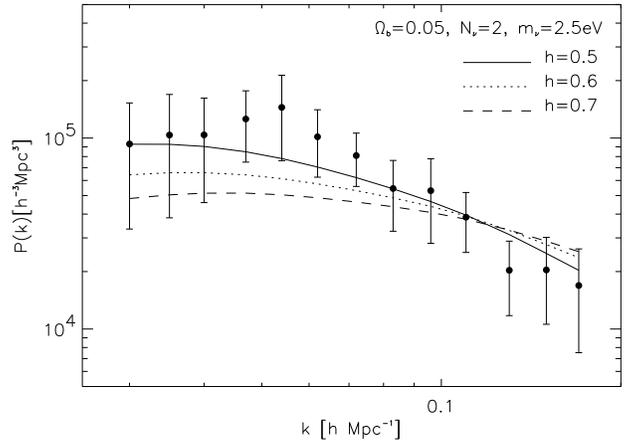} 
\caption{The rich cluster power spectrum for MDM models 
with a varying $h$ ($m_{\nu}$, $N_{\nu}$ and $\Omega_{b}$ are fixed). 
The filed circles are the same as in Fig.1. 
The best-fit biasing parameters for models with $h=$0.5, 0.6 and 0.7 are 
$b_{cl}=$3.2, 2.5, 2.1 accordingly.} 
\end{figure} 
 
Similarly, one shall calculate rich cluster power spectra for 
different $\Omega_{b}$ 
when the rest of the parameters are fixed. The results for $\Omega_{b}=$0.05, 0.1, 0.15, 
0.2, 0.25, 0.3 are presented in Fig.4. The corresponding $\sigma_{8}$'s are 
following 0.71, 0.64, 0.58, 0.53, 0.48, 0.44, the characteristics of 
deviations of the predicted values from the observable ones $\chi^2$ for these models 
are 5.72, 4.28, 3.61, 3.70, 4.59, 6.39. The minimum $\chi^2$ is for model 
with $\Omega_{b}=0.15$. 
\begin{figure}[th] 
\epsfxsize=9truecm 
\epsfbox{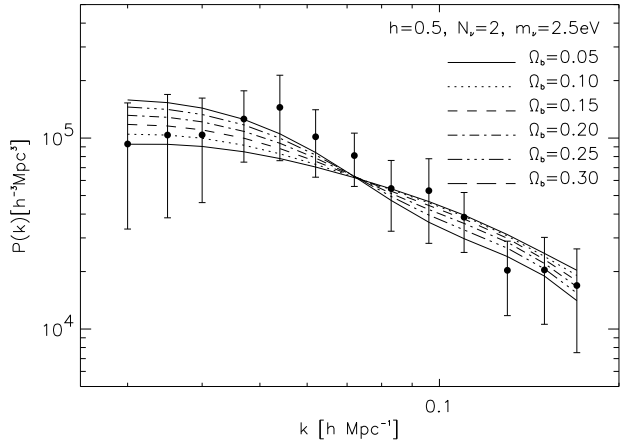} 
\caption{The rich cluster power spectrum  for MDM models 
with a varying $\Omega_{b}$ 
($m_{\nu}$, $N_{\nu}$ and $h$ are fixed). 
The filed circles are the same as in Fig.1. 
The best-fit biasing parameters for models with $\Omega_{b}=$0.05, 0.1, 0.15, 
0.2, 0.25 and 0.3 are $b_{cl}=$3.2, 3.4, 3.7, 3.9, 4.1 and 4.3 accordingly.} 
\end{figure} 
 
As we see the theoretically predicted values of the chosen data are sensitive 
to search parameters $m_{\nu}$, $N_{\nu}$, $\Omega_{b}$ and $h$. 
It is interesting now where the global minimum 
of $\chi^2$ in space of these parameters is when all or a part of them are free. 
 
\section{Results} 
 
The searching of $m_{\nu}$, $N_{\nu}$, $\Omega_{b}$ and $h$ 
by $\chi^2$ Levenberg-Marquardt minimization method can be realized in 
the following way. We shall put $m_{\nu}$, $\Omega_{b}$, $h$ and $b_{cl}$ 
or part of them free and find the minimum of $\chi^2$ for $N_{\nu}$=1, 2, 3 
in a series. 
The lowest value from them will be suggested as minimum of $\chi^2$ for 
each set of free parameters. This is because the $N_{\nu}$ possesses the 
discrete value. 
 
The key point is narrowing the range of search parameter values. 
The analytical approximation of the MDM power spectra used here is 
accurate enough in the following range of parameters: $0.3\le h \le 0.7$, 
$\Omega_{\nu}\le 0.5$, $\Omega_b\le 0.3$, $N_{\nu}\le 3$ (\cite{nov98}). 
By the upper and lower boundaries of $h$, $\Omega_{\nu}$ and $\Omega_b$ 
availability of the used analytical approximation we admeasure the range of 
search values of these parameters. We make these boundaries as 'mirror 
walls'.

\subsection{All parameters are free} 
 
\begin{figure}[th] 
\epsfxsize=9truecm 
\epsfbox{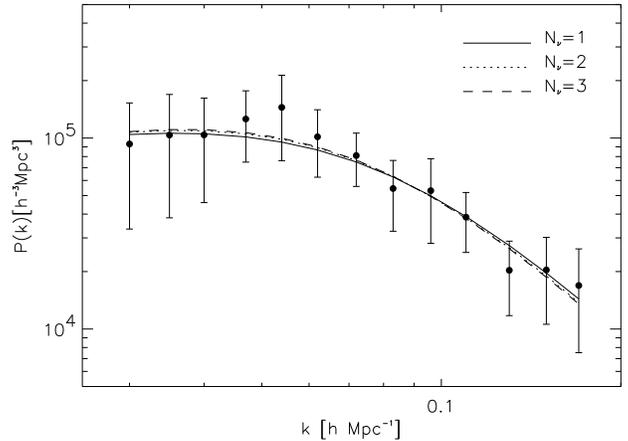} 
\caption{The rich cluster power spectra of MDM models with best-fit 
parameters $\Omega_{\nu}$, $\Omega_{b}$, $h$ and $b_{cl}$ for 1, 2 and 3 
sorts of massive neutrinos (Table 2). 
The filed circles are experimental Abell-ACO power spectrum by Retzlaff 
et al. 1997.} 
\end{figure} 
 
The minima of $\chi^2$ in a 4-dimensional space of parameters 
$\Omega_{\nu}$, $\Omega_b$, $h$ and $b_{cl}$ 
for models with 1, 2 and 3 species of 
massive neutrinos are achieved for the set of parameters presented 
in  Table 2. The spectra for them are shown in Fig.5 and $\sigma_8$'s are 
presented in the Table 2. (The accuracy of analytical approximation of MDM 
spectra is better than 5\%). 

As we can see $\chi^2$ is  few times lower than the formal degree of freedom,
$d=n-m$, where $n$ is the number of data points, $m$ is the number  of free
parameters. The reason is that not all the points of the Abell-ACO power spectrum
presented in Table 1 are independent. The numerical experiment has shown
that the minimal number of points  which determine the same MDM parameters is
$\approx 7$ (odd points of $P_A(k_i)$ in Table 1, for example). Indeed, such a
spectrum can be described by amplitude and inclination at small and
large scale ranges and the second order curve at the peak (or
maximum) range. It means that real $d\approx 3-4$.

\begin{table}[th] 
\caption{Best-fit parameters of MDM models with 1, 2 and 3 sorts of massive 
neutrinos for Abell-ACO power spectrum by Retzlaff et al. 1997 and mass 
function by Girardi et al. 1998.} 
\def\onerule{\noalign{\medskip\hrule\medskip}} 
\medskip 
\halign{#&\quad #&\quad #&\quad #&\quad #&\quad #&\quad #&\quad #&\quad # 
&\quad #&\quad #&\quad #&\quad #&\quad #&\quad #\cr 
\noalign{\hrule\medskip} 
  $N_{\nu}$ & $\chi^2$ & $\Omega_{\nu}$ ($m_{\nu}$) & $\Omega_b$ & $h$ & $b_{cl}$ & $\sigma_8$ \cr 
\noalign{\hrule\medskip} 
  1 & 2.07 & 0.44 (7.2) & 0.0006 & 0.42 & 3.49 & 0.55 \cr 
  2 & 1.77 & 0.47 (5.5) & 0.0014 & 0.50 & 3.37 & 0.56 \cr 
  3 & 1.66 & 0.47 (4.9) & 0.0021 & 0.58 & 3.29 & 0.57 \cr 
\noalign{\hrule\medskip} 
} 
\end{table} 
 
Therefore, in the 5-dimension space of free parameters 
($\Omega_{\nu}$, $N_{\nu}$, $\Omega_b$, $h$ and $b_{cl}$) 
the global minimum of 
$\chi^2$ is achieved for the MDM model with 3 sorts of massive neutrinos. It has 
the lowest $m_{\nu}$ and the highest $h$ which better matches the data on immediate 
measurements of Hubble constant. However, it is unexpected that the found 
$\Omega_{\nu}$ is so high and $\Omega_b$ is so low. They contradict the data 
on high redshift objects and nucleosynthesis constraint 
($0.007\le \Omega_b h^2\le 0.024$, \cite{tyt96,son97,sct97}) accordingly. 
The MDM models with so high a $\Omega_{\nu}$ ($\sim 0.4-0.5$) also have a problem 
with the galaxy formation, $\sigma_0\sim 1$ for them. 
Let's analyze the cases with additional constraints which can lead us 
out of this difficulty. 
 
\subsection{Coordination with nucleosynthesis constraint} 
 
The increasing of baryon content can decrease this difficulty 
(see \cite{eh4}). We shall fix baryon content by the upper limit which is 
resulted from the nucleosynthesis constraint $\Omega_b h^2=0.024$ and keep up 
the rest parameters as free. The found best-fit parameters are in the Table 3, 
rich power spectrum for the case with 3 sorts of massive neutrino is shown 
in  Fig.6 (dotted line). The spectra for the cases with 1 and 2 sorts are 
close to this one. 
\begin{table}[th] 
\caption{Best-fit parameters of MDM models with 1, 2 and 3 sorts of massive 
neutrinos for Abell-ACO power spectrum by Retzlaff et al. 1997 and mass
function by  Girardi et al. 1998 when baryon content is fixed by
nucleosynthesis  constraint ($\Omega_b h^2=0.024$).} 
\def\onerule{\noalign{\medskip\hrule\medskip}} 
\medskip 
\halign{#&\quad #&\quad #&\quad #&\quad #&\quad #&\quad #&\quad #&\quad # 
&\quad #&\quad #&\quad #&\quad #&\quad #&\quad #\cr 
\noalign{\hrule\medskip} 
  $N_{\nu}$ & $\chi^2$ & $\Omega_{\nu}$ ($m_{\nu}$) & $\Omega_b$ & $h$ & $b_{cl}$ & $\sigma_8$ \cr 
\noalign{\hrule\medskip} 
  1 & 2.58 & 0.42 (8.3) & 0.12 & 0.46 & 3.56 & 0.54 \cr 
  2 & 2.02 & 0.46 (6.5) & 0.08 & 0.55 & 3.37 & 0.56 \cr 
  3 & 1.82 & 0.48 (5.7) & 0.06 & 0.62 & 3.27 & 0.57 \cr 
\noalign{\hrule\medskip} 
} 
\end{table} 
As we can see $\Omega_{\nu}$ increases when $\Omega_{b}$ decreases and 
the minima of $\chi^2$  are achieved at high $\Omega_{\nu}$ 
again. But they are quite close to the corresponding minima from the previous table 
($\Delta \chi^2 <1$). 
 
\begin{figure}[th] 
\epsfxsize=9truecm 
\epsfbox{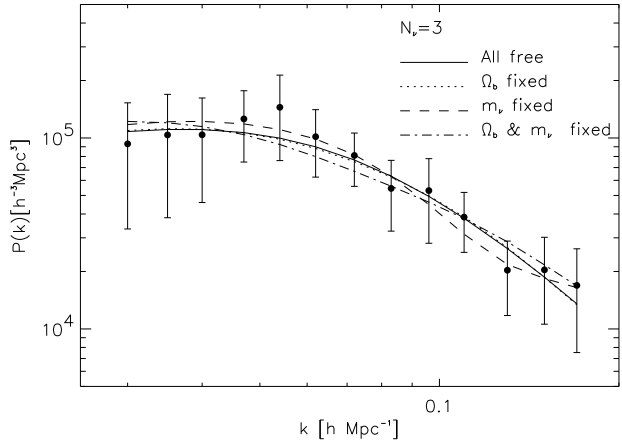} 
\caption{The rich cluster power spectrum of MDM models with 3 sorts of 
massive neutrinos and best-fit parameters for the cases when all 
parameters are free (solid line), when baryon content $\Omega_b$ is fixed by 
nucleosynthesis constraint (dotted line), when mass of neutrino 
$m_{\nu}=2.5eV$ is fixed (dashed line) and when both $\Omega_b$ and $m_{\nu}$ 
are fixed (dashed dotted line). The filed circles are the same as in Fig.5.} 
\end{figure} 
 
\subsection{When the mass of neutrino is known}

An interesting question ensuing from last two items is: which best-fit 
values of $\Omega_b$ and $h$  can be obtained from these data on the Abell-ACO 
power spectrum and mass function in the case 
when mass of neutrino is determined by any physical or astrophysical 
experiments and is known. Let's assume that $m_{\nu}$ is fixed but the number of
species $N_{\nu}$ is unknown. We fix $\Omega_{\nu}$ by relation (4) and 
the rest of parameters leave free. The search in such an approach was unsuccessful 
because it halted in the upper limit of $\Omega_b$=0.3. When this 'mirror wall' 
was removed the solutions were found but with extremely high content of 
baryons for which an accuracy of analytical approximation for MDM spectra 
is worse ($\sim 15-20\%$). Results for $m_{\nu}$=2.5eV and 3eV are presented 
in Table 4. The rich cluster power spectrum for $m_{\nu}=2.5eV$ and $N_{\nu}=3$ 
is shown in Fig.6 (dashed line). The spectra for 1 and 2 sorts are close to this one. 
 
\begin{table}[th] 
\caption{Best-fit parameters of MDM models with 1, 2 and 3 sorts of massive 
neutrinos for Abell-ACO power spectrum by Retzlaff et al. 1997 and mass
function by  Girardi et al. 1998 when neutrino mass is fixed ($m_{\nu}$=2.5
and 3.0eV,  $\Omega_{\nu}=m_{\nu}N_{\nu}/93h^2$).} 
\def\onerule{\noalign{\medskip\hrule\medskip}} 
\medskip 
\halign{#&\quad #&\quad #&\quad #&\quad #&\quad #&\quad #&\quad #&\quad # 
&\quad #&\quad #&\quad #&\quad #&\quad #&\quad #\cr 
\noalign{\hrule\medskip} 
  $N_{\nu}$ & $\chi^2$ & $\Omega_{\nu}$ ($m_{\nu}$) & $\Omega_b$ & $h$ & $b_{cl}$ & $\sigma_8$ \cr 
\noalign{\hrule\medskip} 
  1 & 1.53 & 0.05 (2.5) & 0.47 & 0.75 & 3.22 & 0.65 \cr 
  2 & 1.44 & 0.09 (2.5) & 0.45 & 0.79 & 3.21 & 0.65 \cr 
  3 & 1.39 & 0.12 (2.5) & 0.43 & 0.82 & 3.20 & 0.65 \cr 
  1 & 1.51 & 0.06 (3.0) & 0.47 & 0.75 & 3.21 & 0.65 \cr 
  2 & 1.42 & 0.10 (3.0) & 0.44 & 0.79 & 3.20 & 0.65 \cr 
  3 & 1.37 & 0.14 (3.0) & 0.42 & 0.83 & 3.19 & 0.65 \cr 
\noalign{\hrule\medskip} 
} 
\end{table} 
The $\chi^2$'s in all cases here 
are lower than in Table 2 because the performance of analytical approximation of MDM spectra 
for so high a $\Omega_b$ and $h$ is essentially worse than in the allowance range. 
Therefore we can not conclude that the global minimum of $\chi^2$ in the 
4-dimension space of parameters $m_{\nu}$, $N_{\nu}$, $\Omega_b$ and $h$ 
is in the range of high $\Omega_b$ and $h$. It is in point with the parameters 
which are in the last row of Table 2. But we certainly conclude that when 
$m_{\nu}\sim 2-3eV$ the minimum is absent in the range of $\Omega_b\le 0.3$, 
$0.3\le h\le 0.7$. Therefore the Abell-ACO power spectrum and mass function 
among the MDM models with $m_{\nu}\le 4eV$ and $N_{\nu}\le 3$ prefer 
$\Omega_{b}> 0.3$ and $h\sim 0.8$ that agrees well with the results by 
\cite{eh4}. 
 
\subsection{$\Omega_{b}$ and $m_{\nu}$ are fixed} 
 
One can look now which $h$ is preferable by Abell-ACO power spectrum 
and mass function when neutrino mass and baryon 
content are fixed by the other observable constraints or theoretical arguments. 
Let's put that $m_{\nu}=2.5eV$ ($\Omega_{\nu}=m_{\nu}N_{\nu}/93h^2$) and 
$\Omega_b=0.024/h^2$ is fixed by the upper limit of nucleosynthesis constraint. 
Only $h$ and $b_{cl}$ are free parameters. Their best-fit values found 
for 1, 2 and 3 sorts of massive neutrino are presented in the Table 5. The 
rich cluster power spectrum for $N_{\nu}=3$ MDM model with those parameters 
is shown in Fig.6 (dashed dotted line). 
The spectra for 1 and 2 sorts are close to this one. 
 
\begin{table}[th] 
\caption{Best-fit parameters of MDM models with 1, 2 and 3 sorts of massive 
neutrinos for Abell-ACO power spectrum by Retzlaff et al. 1997 and mass
function by  Girardi et al. 1998 when baryon content and neutrino mass are
fixed:  $\Omega_b=0.024/h^2$, $m_{\nu}$=2.5eV
($\Omega_{\nu}=m_{\nu}N_{\nu}/93h^2$).} 
\def\onerule{\noalign{\medskip\hrule\medskip}}  \medskip 
\halign{#&\quad #&\quad #&\quad #&\quad #&\quad #&\quad #&\quad #&\quad # 
&\quad #&\quad #&\quad #&\quad #&\quad #&\quad #\cr 
\noalign{\hrule\medskip} 
  $N_{\nu}$ & $\chi^2$ & $\Omega_{\nu}$ ($m_{\nu}$) & $\Omega_b$ & $h$ & $b_{cl}$ & $\sigma_8$ \cr 
\noalign{\hrule\medskip} 
  1 & 4.20 & 0.16 (2.5) & 0.15 & 0.41 & 3.90 & 0.55 \cr 
  2 & 3.31 & 0.24 (2.5) & 0.11 & 0.47 & 3.76 & 0.56 \cr 
  3 & 2.85 & 0.29 (2.5) & 0.09 & 0.52 & 3.69 & 0.56 \cr 
\noalign{\hrule\medskip} 
} 
\end{table} 
 
As we see in the MDM model with 3 sorts of 2.5eV neutrinos the best-fit value 
of $h$ and $\sigma_8$ are closer to the corresponding observable data 
than in models with 1 or 2 sorts. 
 
\section{Discussion} 
 
Rich cluster power spectra of models with the best fit parameters are 
within the error bars of the corresponding experimental data (Fig.5-6). 
But none of  them explains the peak at $k\approx 0.05h/Mpc$ 
that corresponds to the linear scale  $\approx 120\Mpc$. 
It has excess power at $\sim 50\%$ in comparison with the best-fit model 
and $\sim 30\%$ in comparison with the high-$\Omega_b$ one. It is more 
prominent yet in the data by \cite{ein97}. Apparently, it is a 
real feature of the power spectrum. The necessity of a similar feature in the power 
spectrum was argued earlier by the explanation of Great Attractor phenomenon 
(\cite{hln95,nov96}). A sample of the Abell-ACO clusters of galaxies used 
by \cite{ret97} is 
placed in $60^o$ double-cone with the axis pointing towards the Milky Way pole. 
The Great Attractor, on the contrary, placed in the plane of our galaxy. 
Therefore, they are an independent experimental demonstration of the reality 
of those peak. Other important arguments for its validity come from 
pencil-beam redshift survey by \cite{bro90} and from 2-dimensional power 
spectrum of the Las Campanas Redshift Survey (\cite{lan96}). The angular 
correlations in the APM survey (\cite{gaz97}) and high-redshift absorption 
lines in quasar spectrum (\cite{qua96}) also show similar features at these 
scales. It was shown also by \cite{aeg97} that this $\sim 120\Mpc$ peak 
well agree with Saskatoon data on the $\Delta T/T$ power spectrum. Therefore, 
the data used here on rich cluster power spectrum are based on the surveys 
which represent a fair sample of $\sim 120\Mpc$ structures and that peak 
is significant despite the large error bars of experimental data.

Obviously, that turnabout to open ($\Omega_0<1$) models or flat with 
cosmological term ($\Omega_0+\Omega_{\Lambda}=1$) does not improve the 
situation with the explanation of that peak in our approach. It is because
the maximum of power spectra in those models is shifted to larger scales 
in comparison with matter dominated flat models analyzed here. 
Explaning of it by baryonic acoustic oscillations calls for extremely high content
of baryons that disagree with nucleosynthesis constraint (see \cite{eh4}). 
Therefore we face a necessity to consider models with a built-in
scale in the primordial power spectrum again. 
 
Let's determine the parameters of this peak. The comparison of 
rich cluster power spectrum predicted  by the MDM model with the best-fit
parameters (Table 2)  with the observable one
showed that the peak has approximately the Gaussian form. Therefore
we approximate it by the function
$p(k)=1+a_{p}exp(2(k_{p}-k)^2/w_{p}^2)$, where $a_p$, $k_p$ and $w_p$ are 
amplitude, center and width of the peak accordingly. We set the power spectrum
in the form of
$P_{MDM+p}(k)=P_{MDM}(k;\Omega_{b},h,m_{\nu},N_{\nu})p(k;a_p,k_p,w_p)$, 
and repeat previous calculations with additional free parameters 
$a_p$, $k_p$ and $w_p$. 
 
It should seem that this peak causes such high best-fit values
of $\Omega_{\nu}$ or $\Omega_b$ in Tables 2-4. The results of the search
for best-fit parameters in the 8-dimensional space of the MDM+peak model
parameters showed that it is not so, that well agrees with the numerical
results by \cite{ret97}.  The introducing of the peak really
decreases the $\chi^2$ but the MDM model parameters are changed weakly.
It is because they are determined mainly by the inclination of the
Abell-ACO power spectrum after the peak and $\tilde \sigma_8$ as the
most accurate
value of the data set used here. The
models with 3 sorts of massive neutrino are preferable like in the previous 
cases. In Table 9 the best-fit parameters of the MDM models with 3 sorts of
massive  neutrino as well as best-fit parameters of the peak are presented 
for 4 cases: all the MDM parameters were free (1st row), baryon content
$\Omega_b$ was fixed by the upper limit of nucleosynthesis constraint (2),
neutrino mass was 
fixed at $m_{\nu}=2.5eV$ (3), $\Omega_b$ and $m_{\nu}$ were fixed (4). The
$\chi^2$ for them are 0.81, 0.86, 1.11, 1.04 accordingly. In all the cases
except (3) the $\sigma_8=0.6$, in (3) case the $\sigma_8=0.66$. The rich
cluster power spectrum for these cases are shown in Fig.7.

\begin{table}[th] 
\caption{Best-fit parameters of MDM+peak models with 3 sorts of massive 
neutrinos for Abell-ACO power spectrum by Retzlaff et al. 1997 and mass 
function by 
Girardi et al. 1998. The fixed parameters are noted by $ ^{(*)}$ 
($\Omega_b=0.024/h^2$, $m_{\nu}$=2.5eV).} 
\def\onerule{\noalign{\medskip\hrule\medskip}} 
\medskip 
\halign{#&\quad #&\quad #&\quad #&\quad #&\quad #&\quad #&\quad #&\quad # 
&\quad #&\quad #&\quad #&\quad #&\quad #&\quad #\cr 
\noalign{\hrule\medskip} 
  $m_{\nu}$ & $\Omega_b$ & $h$ & $b_{cl}$ & $k_p$ & $a_p$ & $w_p$ \cr 
\noalign{\hrule\medskip} 
4.6 & 0.01  & 0.58 & 3.14 & 0.056 & 0.46 & 0.011 \cr 
5.0 & $0.064^{(*)}$  & 0.61 & 3.11 & 0.056 & 0.47 & 0.012 \cr 
$2.5^{(*)}$ & 0.424  & 0.82 & 3.16 & 0.054 & 0.34 & 0.007 \cr 
$2.5^{(*)}$ & $0.084^{(*)}$  & 0.53 & 3.33 & 0.060 & 0.63 & 0.013 \cr 
\noalign{\hrule\medskip} 
} 
\end{table} 
 
\begin{figure}[th] 
\epsfxsize=9truecm 
\epsfbox{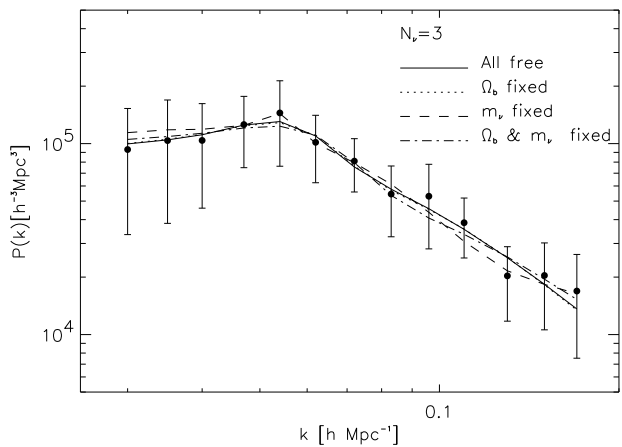} 
\caption{The rich cluster power spectrum of MDM+peak models with 
3 sorts of massive neutrino and best-fit parameters from Table 6. 
The filed circles are the same as in Fig.5.} 
\end{figure}

The introducing of such a peak increases the predicted bulk velocities
in a top-hat sphere of the radius R whose r.m.s. values  can be calculated
according to 
$$V_{R}^2={H_0^2\over 2\pi^2}\int_0^{\infty}dkP_{MDM}(k)W^2(kR),$$ 
where $W(kR)$ is the Fourier transform of this sphere. So, for $R=50\Mpc$ 
it increases from  $340km/s$ to $360km/s$ for the best-fit model (3rd row of
Table 2, 1st row of Table 6) and from $330km/s$ to $345km/s$ for a model
with fixed $m_{\nu}$ and $\Omega_b$ (3rd row of Table 5, the last row of Table 6).
The observable value of bulk velocity for  this scale is
 $\tilde V_{50}=375\pm 85km/s$, which follows from
Mark III POTENT results (\cite{kol97}). Therefore, this peak is preferable also
by the data on large
scale peculiar velocity of galaxies and Great Attractor like structures. 
However, the  models with high values of
$\Omega_{\nu}\sim 0.4-0.5$ ($m_{\nu}\sim 4-7eV$), which are best-fit ones
for the Abell-ACO data, have problems with the explanation of
galaxy scale structures and high redshift objects. But models with 
median $\Omega_{\nu}\sim 0.2-0.3$ ($m_{\nu}\sim 2.5$, $N_{\nu}\sim 2-3$) 
are not ruled out by these data ($\Delta \chi^2<1$). On the contrary, the CDM
model with $\Omega_b\le 0.2$ and $h\ge 0.5$ is ruled out by these data at a
high confidence level because for them $\Delta \chi^2\le 15$. 
 
At last it must be noted that primordial spectrum feature 
like this peak is inherent for double inflation models 
(\cite{kof85,kof87}, Kofman $\&$ Pogosyan 1988, Gottloeber et al. 1991,
\cite{pol92}) and inflationary model wherein  an inflation field evolves
through a kink in the potential (\cite{sta92}).  Both classes of these models
were confronted with the observational data on the  Abell-ACO power spectrum by
\cite{les97} and \cite{ret97} accordingly.   
 
\section{Conclusions} 
 
The Abell-ACO power spectrum by \cite{ret97} and mass function  by \cite{gir98}
in the parameter space of the MDM model ($\Omega_0=1$) prefer a region with high
$\Omega_{\nu}$ ($\sim 0.4-0.5$), low $\Omega_b$ ($\le 0.01$) and $h$ 
($\sim 0.4-0.6$). The best-fit parameters are as follows:
$N_{\nu}=3$, $m_{\nu}=4.4eV$, $h=0.56$, $\Omega_b\le 0.01$. Unfortunately, 
experimental uncertainties of the data used here for the determination of
these parameters give no chance
to rule out models with a different set of parameters at a sufficiently high
confidence level. The MDM models with baryon content at the upper limit of
the nucleosynthesis constraint ($\Omega_bh^2=0.024$) do not outstep
$\Delta \chi^2=1$ of best-fit model (see Table 3). 
The high-$\Omega_b$ ($\sim0.4-0.5$) solutions are obtained when neutrino 
mass are fixed and $\le 3eV$. 
 
Introducing artificially into the primordial power spectrum a peak
of Gaussian form decreases the $\chi^2$, increases the bulk motions but 
does not change essentially the best-fit parameters of the MDM models. It means
that determinative for these parameters is mainly inclination of the Abell-ACO power spectrum
at the scales smaller than the scale of the peak position  and $\tilde \sigma_8$ as
the most accurate value of the data set used here.
 
Hereby, the  power spectrum of the Abell-ACO clusters of galaxies and mass
function are a sensitive test for the MDM model parameters.
But more accurate data on  power spectrum of matter density fluctuations
are necessary for more certain determination of cosmological parameters.

{\it Acknowledgments} This work was performed thanks to the
financial support granted by Swiss National Science Foundation
(grant NSF 7IP050163) and DAAD in Germany (Ref.325). The author
also thanks AIP for hospitality and S. Gottloeber for useful discussions.

\end{document}